# A GPU-based Monte Carlo framework for IMRT QA using EPID transit dosimetry

Ning Gao [1], Didi Li [2], Na Liu [2], Yankui Chang [3], Qiang Ren [4], Xi Pei [4], Zhi Wang [2], Xie George Xu [1]

1 School of Nuclear Science and Technology, University of Science and Technology of China, Hefei, China
2 Department of Radiation Oncology, First Affiliated Hospital of Anhui Medical University, Hefei, China
3 School of Medical Imaging, Bengbu Medical College, Bengbu, China.
4 Anhui Wisdom Technology Co., Ltd., Hefei, Anhui, China

**ABSTRACT**

**Background:** EPID transit dosimetry is an important tool for patient-specific quality assurance (PSQA) in radiotherapy. However, its accuracy depends on the model used to predict the transmitted radiation. While Monte Carlo (MC) simulation is the gold standard for this task, its clinical application is limited by long calculation time.

**Purpose:** We presented a GPU-based MC framework, ARCHER-EPID, specifically designed for EPID transit dosimetry, with improving accuracy and efficiency.

**Methods:** A comprehensive MC framework was developed to perform full radiation transport simulations through three distinct zones: a detailed linear accelerator head model, a CT-based patient/phantom geometry, and a realistic, multi-layered EPID model. To convert the simulated absorbed dose to a realistic detector signal, a dose-response correction model was implemented. The framework was validated by comparing simulations against experimental measurements for 25 IMRT fields delivered to both a solid water phantom and a anthropomorphic phantom. Agreement was quantified using Gamma analysis.

**Results:** The GPU-accelerated ARCHER-EPID framework can complete the simulation for a complex IMRT field in about 90 seconds. A 2D correction factor lookup table is generated by parameterizing radiological thickness and effective field size to account for the EPID's energy-dependent response. The data revealed that for small fields, beam hardening is the dominant effect, while for large fields, the contribution from patient-generated scatter overwhelms this effect. The average 2D gamma passing rates (3%/3 mm criteria) between simulation and measurements are 98.43% for the solid water phantom and 97.86% for the anthropomorphic phantom, respectively. Visual comparison of the images and dose profiles between simulation and measurements show a high degree of agreement.

**Conclusions:** We have successfully developed and validated a GPU-based MC framework that provides gold-standard accuracy for EPID transit dosimetry in radiotherapy. The results demonstrate that our proposed method has potential for routine application in PSQA.

# 1 INTRODUCTION

With the implementation of accurate treatment methods, including Intensity-modulated radiotherapy (IMRT), the electronic portal imaging device (EPID) has become an important tool for patient-specific quality assurance (PSQA) in modern radiotherapy[1,2]. Its high resolution and integration with the linac make it suitable for complex IMRT dose distributions, as emphasized by AAPM TG-307 report[3]. Comparing with pre-treatment verification, in vivo dosimetry——performed with the patient in the treatment position—— is highly sensitive to patient-relevant deviations, such as tumor regression, weight loss, and organ motion[2]. EPID-based in vivo dosimetry, known as EPID transit dosimetry, verifies the accuracy of the dose delivery by comparing the measured transmission image against a predicted image[4]. The challenge of this method lies in developing a dosimetry model that could predict the EPID image with both high accuracy and efficiency.

Traditional methods used to rely on empirical or semi-empirical algorithms[5-8]. However, their accuracy is fundamentally challenged by the complex physics of EPID. Due to the high-atom-number components in their construction, EPID exhibits a strong energy-dependent detector response, particularly to low-energy photons[5,9]. More importantly, the presence of the patient introduces a large amount of scattered radiation that varies with patient thickness and radiation field size, causing the "dose to EPID" to differ substantially from the "dose in water". To address these issues, indirect dose prediction methods have been developed to generate a final dose image by convolving a predicted fluence with pre-determined scatter kernels[5]. Another common approach leverages a treatment planning system as a dose calculation engine[10-13]. However, by modeling EPID as a water-equivalent phantom, this method fails to consider the true signal response of the detector's physical materials.

Compared with these indirect methods, Monte Carlo (MC) simulation is recognized as the gold standard in dose calculation for its ability to model the radiation transport process. MC-based methods can inherently account for complex scatter photons produced inside the patient[14-16]. Previous works have demonstrated the accuracy of MC for predicting EPID dose images in both transit and non-transit dosimetry. Unfortunately, the primary drawback of this method has been its computational cost. To achieve an acceptable level of statistical uncertainty, a very large number of particle histories must be simulated, consuming tens of hours for a single complex IMRT field on traditional CPU-based systems. While various strategies have been proposed to accelerate the simulation process, the calculation time remains on the order of hours, making it impractical in routine clinical QA[17].

The emergence of Graphics Processing Unit (GPU) accelerated computing has presented a powerful solution to reduce the calculation time during MC simulations. The technology has been successfully employed to accelerate MC simulations in several areas of medical physics, including dose calculations[18] and imaging simulations[19]. However, its application to EPID transit dosimetry remains to be explored.

In this study, we develop and validate a GPU-based MC framework, ARCHER-EPID, specifically designed for EPID transit dosimetry. We integrate a comprehensive model that performs detailed radiation transport simulations in a linac treatment head, patient-specific anatomy, and a realistic EPID structure. In addition, we employ a simple calibration method to calibrate the EPID dose image to the measured EPID image, accounting for the dosimetric response of EPID. The accuracy and efficiency of this framework are then validated through experimental measurements on both homogeneous and anthropomorphic phantoms, demonstrating the potential of IMRT in vivo dosimetry.

## 2 METHODS

### 2.1 GPU-based MC framework: ARCHER-EPID

The framework proposed in this study is based on ARCHER, which is an MC computational engine designed for heterogeneous CPU/GPU environments[20]. ARCHER has been validated for a variety of areas in medical physics, involving radiotherapy, CT imaging, and shielding design[20-22]. The core of our method is built upon ARCHER-RT, specifically designed for photon radiotherapy[21]. This study extends the capabilities of

ARCHER-RT into the domain of EPID-based QA. As illustrated in Figure 1, the ARCHER-EPID module performs a full simulation of particle transport through three primary geometric zones: (1) linac treatment head, (2) patient or phantom, and (3) EPID.

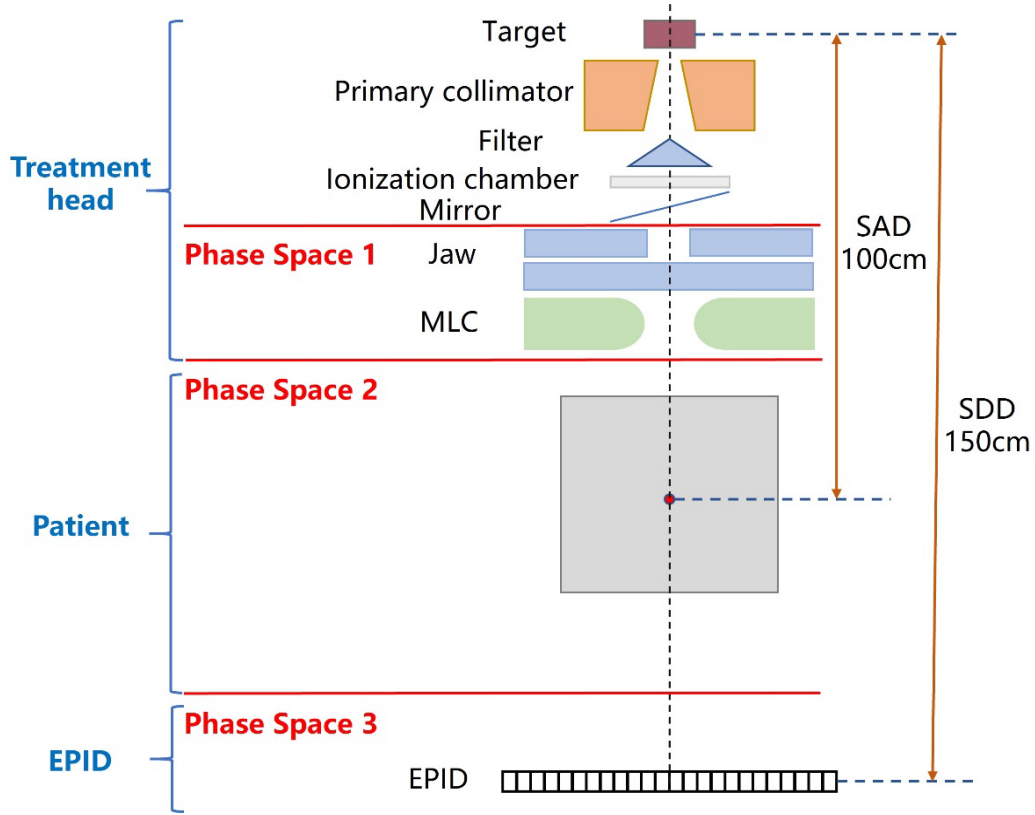

FIGURE 1 Schematic diagram of the ARCHER-EPID Monte Carlo simulation workflow. The process is divided into three distinct zones: (1) particle transport within the linac treatment head, (2) transport through the patient or phantom, and (3) dose scoring within the EPID model.

2.1.1 Simulation inside the linac treatment head

A model of the Varian (Varian Medical Systems, Palo Alto, CA, USA) VitalBeam linac treatment head was developed for the radiation transport simulations. To enhance computational efficiency, a two-stage process was implemented that separates patient-independent and patient-specific components. The patient-independent components, involving target, primary collimator, and so on, were pre-simulated once to generate Phase Space 1 (as shown in Figure 1), which is stored and serves as a reusable particle source. In the second stage, particles were sampled from Phase Space 1 to simulate their transport through the patient-specific components: the jaws and the multi-leaf collimator (MLC). The geometric positioning and aperture shapes of these components were extracted from the patient's RTPLAN DICOM file. Following transport through the jaw and MLC models, a second phase space, Phase Space 2, was generated and directly streamed to the patient simulation. Throughout this study, the term "number of particle histories" refers to the total number of particles used in Phase Space 2. The accuracy of this part has been validated by previous studies[23].

### 2.1.2 Simulation inside the patient

Particle transport inside the patient or phantom was developed based on DPM MC code (), which is integrated into the ARCHER-EPID framework[21]. The transport process simulated all primary and secondary particle interactions until they either fell below a pre-defined energy cutoff or exited the simulation geometry. The energy cutoffs for particle transport were set to 10 keV for photons and 200 keV for electrons. For photon transport, the photoelectric effect, Compton scattering, and pair production were modeled. Electron transport was handled using a Class II condensed history method, which modeled discrete hard collisions and used a continuous slowing-down approximation for soft collisions. A DICOM parser implemented within ARCHER-EPID could automatically process patient data, including CT, RT DICOM files. The patient or phantom anatomy was represented as a three-dimensional density matrix derived from CT images using a HU-to-density curve[24].

### 2.1.3 Simulation inside EPID

In terms of energy-dependent property, EPID was modeled by a realistic, multi-layered structure accounting for heterogeneity in the actual composition of EPID construction. The specific thickness, mass density, and elemental composition for each layer—including the copper buildup plate, reflective layer, $Gd_2O_2S$ phosphor scintillator, and protective layers—are considered[23]. The native 0.336 mm pixel pitch of the AS1200 detector is too fine for practical MC simulation, as achieving adequate statistical certainty across such a large number of voxels would be computationally prohibitive, even with

GPU acceleration. To balance statistical uncertainty with computational efficiency, a preliminary study was conducted to determine an optimal simulation resolution, leading to the selection of a 2.5 mm × 2.5 mm pixel size for dose scoring in this work. This resolution was chosen to improve computational efficiency while maintaining sufficient accuracy in the calculated dose distributions, as supported by a previous study[23]. To couple the patient and EPID simulations, an additional scoring plane, Phase Space 3, is generated to store all particles exiting the patient. This includes primary photons as well as scattered photons and electrons generated within the patient's anatomy. While doses are simulated in a realistic EPID model, complex optical transport and signal processing are not involved. Therefore, a further correction step is required to relate the simulated dose to the measured imager response, as will be described in the following section 2.2

**2.2 EPID dose-response correction model**

The purpose of the dose-response correction is to convert the MC-simulated EPID dose image into a signal that accurately represents the real-world detector response, enabling a direct comparison between simulated and measured images. As the EPID's response is primarily influenced by the incident photon energy spectrum, which is in turn affected by the patient/phantom thickness and the radiation field size, we introduced a two-dimensional correction factor, $k(t,s)$, where $t$ represents the radiological thickness and $s$ represents the effective field size. This correction factor is defined as the ratio of the measured EPID signal to the corresponding MC-simulated dose under identical, well-defined conditions. To generate a comprehensive lookup table for $k(t,s)$, a series of measurements and simulations were performed using a solid water phantom. The parameter space included different phantom thicknesses ($t$ = 0, 5, 10, 15, and 20 cm) and different square field sizes ($s$ = 5 cm × 5 cm, 10 cm × 10 cm, 15 cm × 15 cm, and 20 cm × 20 cm).

For application to a clinical case, the required inputs are determined as follows:

1. Radiological Thickness ($t$): A per-pixel radiological thickness map is generated by performing a ray-tracing algorithm from the radiation source through the patient's

CT dataset to the EPID plane. This was implemented using the open-source TIGRE library in Python[25].

2. Effective Field Size ($s$): For complex, irregularly shaped IMRT fields delivered by MLCs, a single field size is not well-defined. To simplify this, the irregular aperture was first approximated by its minimum bounding rectangle, with side lengths $a$ and $b$. The equivalent square field size, $s$, was then calculated using the formula:

$$s = \frac{a * b}{2 * (a + b)} \tag{1}$$

## 2.3 Experimental validation

To validate the clinical efficacy of the method proposed in this study, a series of experimental measurements were performed in both homogeneous and heterogeneous media. All measurements were performed at the Department of Radiation Oncology, The First Affiliated Hospital of Anhui Medical University. A total of 25 IMRT fields from 5 clinically-approved patient plans were selected for validation. All simulations were performed using $1 \times 10^9$ particle histories to ensure the statistical uncertainty was below 1%, and were run on an NVIDIA GeForce RTX 3090 (24GB) GPU. The corresponding experimental fields were delivered using a Varian VitalBeam linear accelerator equipped with an AS1200 EPID.

### 2.3.1 Homogeneous Phantom

To evaluate the algorithm's performance in a uniform medium, a solid water phantom with a thickness of 20 cm was placed on the treatment couch. The IMRT plans were delivered with the gantry fixed at 0°, using a source-to-surface distance (SSD) of 90 cm and a source-to-imager distance (SID) of 150 cm. The transmitted EPID images were acquired simultaneously during each delivery.

### 2.3.2 Anthropomorphic phantom

To test the algorithm's accuracy in a more clinically realistic scenario, a CIRS E2E®

SBRT 036S anthropomorphic phantom was used. This phantom is designed to approximate the anatomy of a human thorax, including complex structures for the spine, ribs, and lungs, all composed of tissue-equivalent epoxy resins suitable for MV-level dosimetric validation. The phantom was positioned on the treatment couch with its geometric center aligned to the machine isocenter, corresponding to a source-to-axis distance (SAD) of 100 cm. The EPID was positioned at an SID of 150 cm. For each of the 25 test fields, the gantry was rotated to the angle specified in the original treatment plan to assess the model's performance across different beam orientations.

### 2.3.3 Data post-processing and analysis

The dose was scored in the EPID model with a pixel size of 2.5 mm × 2.5 mm, resulting in a 160 × 160 pixel image. In contrast, the measured EPID images acquired from the Varian VitalBeam linac have a native resolution of 1190 × 1190 pixels. To enable a direct, pixel-to-pixel comparison, the lower-resolution simulated image was up-sampled to match the dimensions of the measured image. The agreement between the final, corrected MC-predicted EPID image and the corresponding experimentally measured EPID image was quantified using the Gamma Index. Gamma comparisons were conducted using global passing criteria of 3%/3 mm, 3%/2 mm, and 2%/2 mm, all with a 10% low-dose threshold.

## 3 Results

### 3.1 Dose-response correction model

To convert the simulated dose to a signal comparable with the measured EPID image, a dose-response correction factor, *k(t,s)*, was developed. This factor, which is a function of phantom thickness (*t*) and field size (*s*), is defined as:

$$k(t,s) = \frac{M(t,s)}{S(t,s)} \qquad (2)$$

Where *M(t,s)* is the average measured signal within a 1 cm × 1 cm central region of the EPID image, and *S(t,s)* is the corresponding average simulated dose in the same region. The empirically determined values for *k(t,s)* are presented in Table 1. For application to a clinical simulation, the final correction factor for any given pixel is obtained via bilinear interpolation from this table, enabling the transformation of the simulated dose value into the expected measured signal.

Table 1 Empirically determined dose-response correction factors, k(t,s), as a function of solid water phantom thickness and square field size.

| Field Size | 0 cm | 5 cm | 10 cm | 15 cm | 20 cm |
|---|---|---|---|---|---|
| 5 cm × 5 cm | 0.999 | 0.974 | 0.960 | 0.949 | 0.949 |
| 10 cm × 10 cm | 1.001 | 0.975 | 0.967 | 0.964 | 0.953 |
| 15 cm × 15 cm | 0.995 | 0.982 | 0.983 | 0.996 | 0.986 |
| 20 cm × 20 cm | 0.991 | 0.997 | 1.015 | 1.034 | 1.043 |

The data in Table 1 demonstrates a complex relationship between the correction factor, phantom thickness, and field size, driven by two competing physical phenomena: beam hardening and patient-generated scatter. For smaller field sizes, the correction factor consistently decreases as phantom thickness increases, which is attributable to beam hardening. It is important to note that the baseline condition (t=0 cm) represents irradiation through the treatment couch alone. With a water-equivalent thickness of approximately 0.7 cm, the Varian treatment couch acts as a beam-hardening filter. With the increase of field size, the contribution of low-energy scatter photons generated within the phantom becomes the dominant effect. Therefore, for larger field sizes, the correction factor increases with thickness. EPID's over-response to low-energy scatter photons causes the measured signal to increase more rapidly than the simulated dose.

### 3.2 Validation on a solid water phantom

To validate the accuracy of the proposed EPID simulation method in a uniform medium, the MC-predicted images (test images) were compared against the experimentally measured images (reference images) using gamma analysis. The distribution of gamma pass rates across all 25 IMRT fields under different analysis criteria is shown in Figure 2.

As expected, the overall gamma pass rates decreased, and the data dispersion increased as the analysis criteria became more stringent. Under the 3%/3 mm criterion, the analysis showed excellent agreement, with a tight distribution and a median pass rate of approximately 98.5% with almost no outliers. When the criteria were tightened to 2%/2 mm, the distribution became wider, and the median pass rate decreased to approximately 92%, indicating a higher sensitivity to small discrepancies.

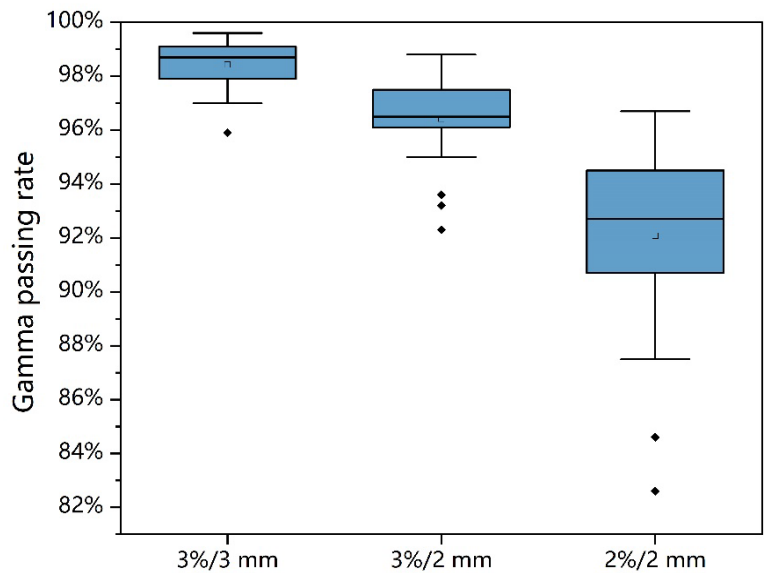

FIGURE 2 Gamma pass rate distributions comparing simulated versus measured EPID images on the solid water phantom. The plots show the results for three global gamma criteria (from left to right): 3%/3 mm, 3%/2 mm, and 2%/2 mm.

A visual comparison of the simulated and measured EPID images further illustrates the high degree of agreement, as shown in a representative field in Figure 3. The overall shape and intensity distribution of the simulated image closely match the measurement. In the high-dose regions, the isodose contours are nearly identical, demonstrating the model's accuracy in clinically significant areas. The primary discrepancies are observed in the low-dose, high-gradient regions at the field edges, where the simulated image appears slightly more blurred. This is an expected consequence of the 2.5 mm simulation resolution, which is slightly coarser than the native resolution of the physical EPID.

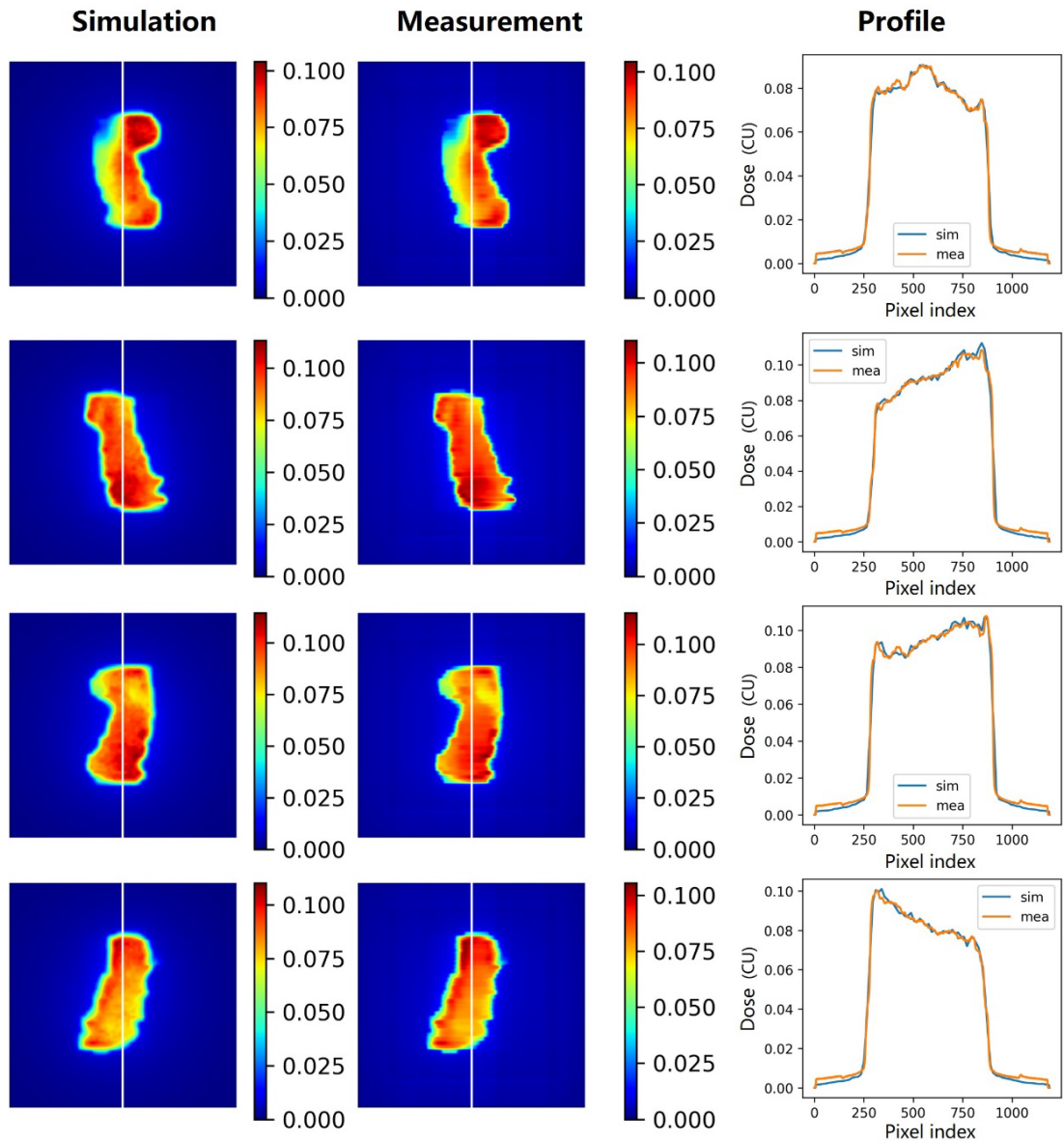

FIGURE 3 Visual comparison of simulated versus measured EPID images for representative IMRT fields on the solid water phantom. Each row corresponds to a single field. The columns show, from left to right: the simulated EPID image, the corresponding measured EPID image, and a central axis line profile comparison. In the profile plots, the blue curve represents the simulated data and the orange curve represents the measured data.

## 3.3 Validation on an anthropomorphic phantom

Figure 4 shows the distribution of gamma pass rates for the 25 IMRT fields under the three analysis criteria on the anthropomorphic phantom. Notably, under the 3%/3 mm criterion, the median gamma pass rate is comparable to that observed for the homogeneous solid water phantom. This result is significant as it demonstrates the robustness of the radiological thickness-based correction method; it indicates that the model effectively accounts for the influence of thickness variations on the EPID response, even in the presence of complex tissue heterogeneities like bone and lung. Consistent with the solid water results, the data distribution is tight and highly consistent at the 3%/3 mm level, with short box-and-whisker ranges. As the criteria become more stringent, the pass rates decrease and the data dispersion increases, reflecting the higher sensitivity of these stricter metrics.

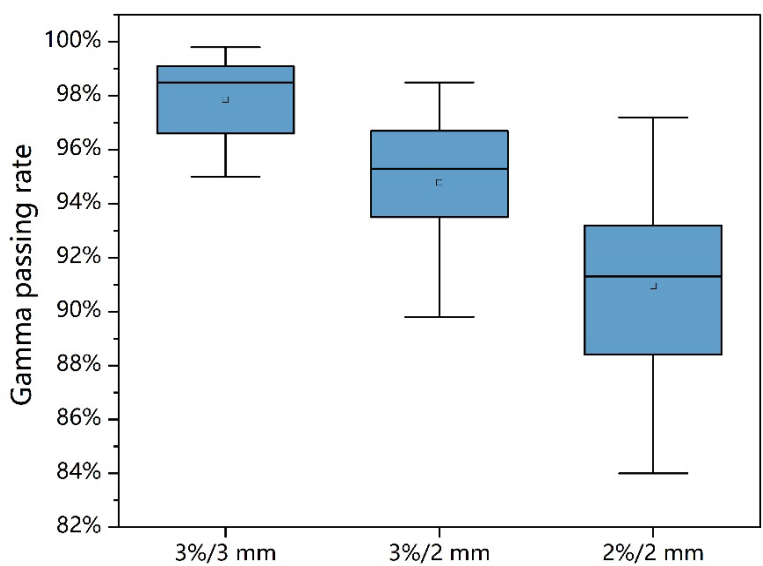

FIGURE 4 Gamma pass rate distributions comparing simulated versus measured EPID images on the anthropomorphic phantom. The plots show the results for three global gamma criteria (from left to right): 3%/3 mm, 3%/2 mm, and 2%/2 mm.

Figure 5 displays a visual comparison of simulated images, measured images, and central-axis dose profiles for five representative IMRT fields delivered through the anthropomorphic phantom. A high degree of qualitative agreement is evident across all fields; the overall shape and intensity patterns of the simulated images closely match the measurements. The corresponding dose profiles are also in excellent agreement, demonstrating a strong match even for larger and more complex field shapes.

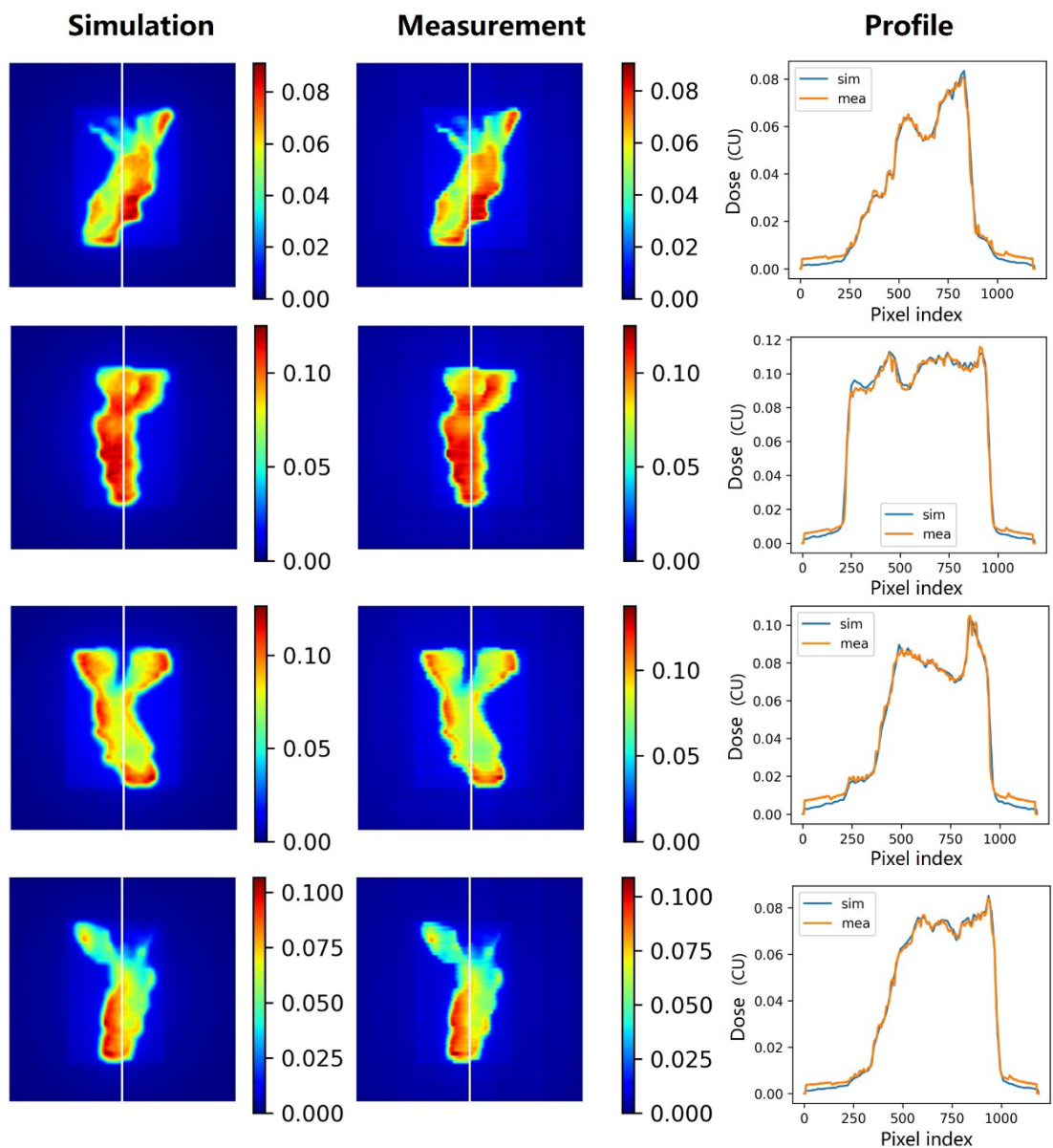

FIGURE 5 Visual comparison of simulated versus measured EPID images for representative IMRT fields on the anthropomorphic phantom. Each row corresponds to a single field. The columns show, from left to right: the simulated EPID image, the corresponding measured EPID image, and a central axis line profile comparison. In the profile plots, the blue curve represents the simulated data and the orange curve represents the measured data.

# 4 DISCUSSION

## 4.1 Accuracy and efficiency of the proposed method

In this study, we developed and validated a GPU-based Monte Carlo framework, ARCHER-EPID, for fast and accurate prediction of EPID transit images. Our results demonstrate that this approach successfully improves MC calculation speed with high accuracy. Studies using conventional MC codes like BEAMnrc/DOSXYZnrc or Geant4 have reported simulation times ranging from 5 hours to as long as 94 hours for a single treatment field[14,26-28]. Even more recent efforts aimed at creating 'fast' MC models, such as the work by Yoon et al., which focused on improving physical accuracy, were still fundamentally limited by their CPU-based architecture and reported calculation times of 1.2 hours[29]. In contrast, our proposed method can predict a complex IMRT field EPID image in approximately 90 seconds.

More importantly, this speed is achieved without compromising dosimetric accuracy. The validation results show average gamma passing rates at 3%/3 mm of 98.43% for the homogeneous solid water phantom and 97.86% for the heterogeneous anthropomorphic phantom. This level of accuracy is significantly higher than that reported for many analytical or TPS-based algorithms, which have shown pass rates of 94.2% and 86.8% respectively, under similar conditions[12]. This demonstrates the dosimetric advantage of a full MC simulation that can accurately model the complex physics of patient scatter and detector response.

While a 90-second calculation time is clinically practical for post-treatment verification, achieving true real-time analysis requires further acceleration. A promising avenue for future work is the integration of deep learning (DL)-based denoising. As demonstrated by Zhang et al. [30] for 3D patient dose calculations, a trained neural network can effectively remove the statistical noise from a low-particle-count simulation to recover an image with the quality of a high-particle simulation, but in a fraction of the time. By combining rapid, low-count MC simulations with DL denoising, it may be possible to reduce prediction times to mere seconds, paving the way for real-time error detection and intervention.

## 4.2 Modularity and generalizability of the method

ARCHER-EPID framework was designed with a modular structure, separating the simulation into three distinct components: the linac treatment head, the patient phantom, and the EPID model. This modularity allows the framework to be generalized beyond the Varian VitalBeam accelerator validated in this study. The patient model is generated from any standard CT dataset, and a new EPID model can be constructed based on the specifications of any detector. Similarly, treatment head models can be incorporated using automated beam modeling methods[31]. With the availability of validated models for a wide range of commercial accelerators, ARCHER-EPID framework can be readily extended to other platforms, making it a broadly applicable solution.

## 4.3 Limitations

One limitation of this study is the simplified dose-response correction model, relying on radiological thickness and equivalent field size. While shown to be effective, a more physics-based model that explicitly incorporates per-pixel spectral information could further improve accuracy. For instance, Zhang et al. employed the PRIMO Monte Carlo program to compute EPID dose images and then used deep learning to convert these dose images into EPID images[16]. In addition, this work focused solely on the validation of static IMRT fields. Extending the framework to handle VMAT deliveries is a critical next step.

## 5 CONCLUSIONS

This study developed a GPU-based MC framework, ARCHER-EPID, specifically for EPID transit dosimetry. The framework is capable of predicting an EPID image for an IMRT field in approximately 90 seconds. In addition, we implemented a dose-response correction model to convert the simulated dose into a realistic detector signal. The final method demonstrated excellent accuracy, with average gamma pass rates (3%/3 mm) of 98.43% and 97.86% for homogeneous and anthropomorphic phantoms, respectively.

These results demonstrate that ARCHER-EPID has the potential to serve as a powerful and efficient tool for PSQA in radiotherapy.

## ACKNOLEDGMENTS

## CONFLICT OF INTEREST STATEMENT